          [2001/04/25 1.1 (PWD)]
\documentclass[a4paper,twocolumn]{esapub} % European paper
\usepackage{natbib}
\usepackage{epsfig}

\title{SPI observations of positron annihilation radiation from the
4th galactic quadrant: sky distribution}
\author{G. Weidenspointner\footnote{ESA Fellow}}
\author{V. Lonjou}
\author{J. Kn{\"o}dlseder}
\author{P. Jean}
\author{M. Allain}
\author{P. von Ballmoos}
\author{M.J.~Harris}
\author{G.K. Skinner}
\author{G. Vedrenne}
\affil{Centre d'{\'E}tude Spatiale des Rayonnements, BP 4346, 31028
Toulouse, France} 
\author{B.J. Teegarden}
\author{N. Gehrels}
\affil{NASA/GSFC, Code 661, Greenbelt, MD 20771, USA}
\author{N. Guessoum}
\affil{American University of Sharjah, Physics Department, Sharjah,
UAE}
\author{V.~Sch{\"o}nfelder}
\affil{Max-Planck-Institut f{\"u}r extraterrestrische Physik, Postfach
1312, D-85741 Garching, Germany}
\author{C.~Chapuis}
\author{Ph.~Durouchoux}
\affil{CEA Saclay, 91191 Gif-sur-Yvette, France}
\author{E. Cisana}
\author{M. Valsesia}
\affil{IASF CNR, 20133 Milan, Italy, and Universita degli studi di
Pavia, 27100 Pavia, Italy}

\begin{document}


\maketitle

%%%%%%%%%%%%%%%%
%   abstract   %
%%%%%%%%%%%%%%%%

\begin{abstract}
During its first year in orbit the INTEGRAL observatory performed deep
exposures of the Galactic Center region and scanning observations of
the Galactic plane. We report on the status of our analysis of the
positron annihilation radiation from the 4$^{th}$ Galactic quadrant with
the spectrometer SPI, focusing on the 
%morphology 
sky distribution of the 511~keV 
%annihilation line radiation
line emission. The analysis methods are described; current constraints
and limits on the Galactic bulge emission and the bulge-to-disk ratio
are presented.
\end{abstract}

%%%%%%%%%%%%%%%%%%%%
%   introduction   %
%%%%%%%%%%%%%%%%%%%%

\section{Introduction}
\label{intro}

The cosmic positron annihilation radiation was first detected through
its hallmark, a gamma-ray line at 511~keV, in balloon observations of
the Galactic Center (GC) region in the 1970s and has been the focus of
intense scrutiny by a large number of balloon and satellite borne
experiments ever since 
%\citep[see reviews by, e.g.,][]{lingenfelter_ramaty89, harris97}.
\citep[e.g.][]{harris97}.
Despite this tremendous effort, the origin of the positrons is still
far from being understood.
A large variety of positron sources and production mechanisms have
been proposed over the years 
%\citep[see e.g.\ discussions in][]{chan_lingenfelter93,
%ramaty_lingenfelter94}
\citep[e.g.][]{chan_lingenfelter93}. Among the more
promising source candidates are the nucleosynthesis products from
supernovae. More recently, hypernovae/GRBs
\citep{casse04} and the annihilation 
%or decay 
of dark matter 
%or relic particles \citep[e.g.][]{boehm04, hooper_wang04} have been
\citep[e.g.][]{boehm04} have been
revisited. Another intriguing candidate source of positrons comprises
compact objects.

Investigations of the 
%morphology 
sky distribution of the annihilation radiation promise
to provide clues for identifying the source(s) of positrons
in our Galaxy, despite the fact that positrons may travel from their
source before annihilating. The annihilation radiation was mapped for
the first time with the OSSE instrument on board the Compton Gamma-Ray
Observatory \citep[e.g.][]{purcell97, milne01}. At least two extended
spatial components, a (possibly dominant) extended bulge and a disk, are
required to account for the observed annihilation emission. 
%The important bulge-to-disk ratio is only poorly constrained.
Early OSSE maps showed tantalizing evidence for a third emission
component (the so-called positive latitude enhancement, PLE) located
about $8^\circ$ north of the 
%Galactic center
GC \citep{purcell97}; however, the final analysis did not corroborate
this \citep{milne01} and its (astrophysical and/or instrumental)
origin remains uncertain \citep{vonballmoos03}.

Modelling of the Galactic distribution of the 511~keV line emission
from OSSE data suggests FWHM values of about $4^\circ - 6^\circ$ when using
Gaussians to describe the bulge component. No significant offset from
the GC
%Galactic center 
was found by \citet{purcell97} or \citet{milne01}. However,
\citet{tueller96}, using OSSE observations with the long axis of the
collimator perpendicular to the Galactic plane to produce a maximum
sensitivity scan in longitude, derived a significant ($\sim 4\sigma$)
offset of the 511~keV line emission. Analyzing data from the wide
field-of-view TGRS instrument, \citet{harris98} obtained a rather
extended distribution (FWHM $\sim 25^\circ$) with no significant
offset from the GC.
%Galactic center. 
%
%Before SPI, OSSE was the only instrument that allowed the
%bulge-to-disk ratio (hereafter B/D) to be constrained.
%%, the ratio of the fluxes from the Galactic bulge and disk.
%This parameter, important for characterizing the Galactic distribution
%of the emission, could only be poorly constrained to a range of 0.2 --
%3.3.  
Before SPI, OSSE was the only instrument that allowed the
bulge-to-disk ratio (hereafter B/D) to be constrained, albeit poorly,
to a range of 0.2 -- 3.3. This parameter is important for
characterizing the Galactic distribution of the emission and -
potentially - of positron sources.
A major uncertainty in the determination of B/D is the possible
existence of an extended (halo) component to the bulge, which would
lead to a large ratio. 
%Unlike B/D, the total 511~keV line flux from
%their model Galaxy is relatively well determined: $(2.1-3.1)
%\times 10^{-3}$~ph~cm$^{-2}$~s$^{-1}$ \citep{milne00}.
Using OSSE, \citet{milne00} could determine the total 511~keV line
flux from their model Galaxy relatively well (unlike B/D): $(2.1-3.1)
\times 10^{-3}$~ph~cm$^{-2}$~s$^{-1}$.

Early SPI results on the 511~keV line emission from the GC
%Galactic center 
region have been reported by \citet{jean03} and
\citet{knoedlseder03}. The Galactic distribution of the emission could be
described with a Gaussian bulge component of about $9^\circ$
FWHM and a $2\sigma$ uncertainty range covering $6^\circ -
18^\circ$. The centroid of the Gaussian was found to be consistent
with the GC
%Galactic center 
at the $2.1\sigma$ level. No evidence for a Galactic disk
component was found. However, flux limits were in agreement with
OSSE measurements. 
%The early data did not allow the reported PLE of 
%\citet{purcell97} to be constrained.

In this paper we report results on the sky distribution of the 511~keV
positron annihilation line radiation from the GC
%Galactic center 
region after one year of observations with the spectrometer SPI on
board ESA's INTEGRAL observatory. Spectroscopic analysis of the
511~keV line, using the same data, is presented by
\citet{lonjou04a}. The results from both of these studies are
presented in the context of previous findings by \citet{jean04}; their
astrophysical implications are discussed by \citet{guessoum04}.

%%%%%%%%%%%%%%%%%%%%%%%%%%%%%%%%%%%%
%   instrument and data analysis   %
%%%%%%%%%%%%%%%%%%%%%%%%%%%%%%%%%%%%

%\section{Instrument and Data Analysis}
\section{Data Analysis}
\label{data_analysis}

%The Spectrometer for INTEGRAL \citep[SPI,][]{vedrenne03} is one of the
%two main instruments on board the INTEGRAL mission.
%%A detailed description of SPI can be found in \citet{vedrenne03}. 
%The detector plane of SPI consists of
%an array of 19 actively cooled high-resolution Ge detectors, which
%cover an energy range of 20--8000~keV at an energy resolution of
%2--8~keV FWHM. SPI employs an active anti-coincidence shield made of
%bismuth germanate, which also defines the aperture of the
%instrument. A coded mask, consisting of tungsten elements and located
%1.7~m above the Ge detectors, allows SPI to image the sky with
%an angular resolution of about $2.5^\circ$. The fully coded
%field-of-view is about $16^\circ$ in diameter.

The data used for the analysis of the sky distribution of the 511~keV
line emission with the spectrometer SPI \citep{vedrenne03} consist of
the spring and autumn Galactic Center Deep Exposures of 2003
(hereafter GCDE), supplemented by Galactic Plane Scan (GPS)
observations of the central Galaxy. Both the GCDE and the GPS
observations are part of the so-called Core Program observations of
the INTEGRAL observatory, which are proprietary to the INTEGRAL
Science Working Team (ISWT) and the instrument teams for one year
\citep{winkler01}. 
%\citep{winkler01}. 
The data comprise the INTEGRAL orbital revolutions
47--66 (March~3 -- April~28) and 97--123 (August~2 -- October~16),
combining 1266 and 1068 pointings with a total exposure time of about
$1.8$ and $2.0 \times 10^6$~s, respectively.
%
%The GCDE consists of three rectangular pointing grids covering
%Galactic longitudes $|l| \le 30^\circ$ and Galactic latitudes $|b| \le
%10^\circ$, and a fourth grid covering the same longitude range but
%$|b| \le 20^\circ$ in latitude. The GPS consists of pointings,
%following a saw-tooth pattern, along the Galactic plane within $|b|
%\le 6.4^\circ$. 
The GCDE consists of rectangular pointing grids covering
Galactic longitudes $|l| \le 30^\circ$ and Galactic latitudes $|b| \le
20^\circ$; the GPS consists of pointings along the Galactic plane within $|b|
\le 6.4^\circ$ \citep[see][ for details]{winkler01}.
%
%Details on the core program observing strategy can be found in
%\citet{winkler01}.

%Only single detector events were used. Before combining the data we
%checked that the two GCDEs, when analyzed individually, give
%consistent results with respect to emission intensity and morphology.

\begin{figure}
\centering
\epsfig{file=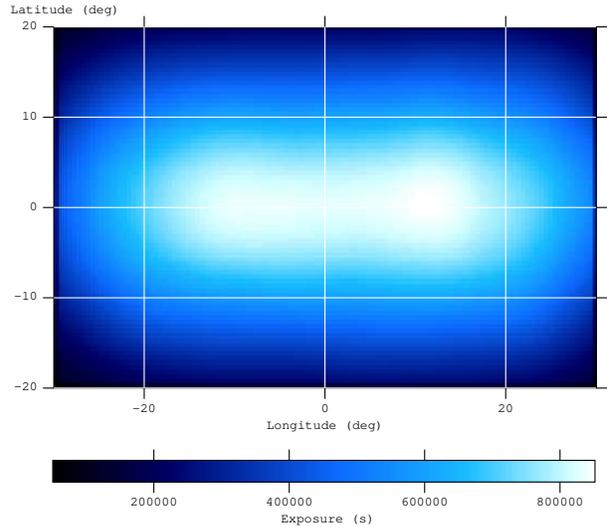,width=8cm,%
bbllx=80pt,bblly=198pt,bburx=535pt,bbury=598pt,clip=}
\caption{The sky exposure resulting from the Core Program observations
used in the analysis \label{exposure_map}}
\end{figure}

Due to data sharing agreements within the ISWT,
%INTEGRAL Science Working Team, 
the results presented
here are limited to the $4^{th}$ Galactic quadrant ($l = 270^\circ -
360^\circ$). However, in accordance with the aforementioned
agreements, data from pointings in the entire GCDE region (as defined
above) have been included in the analysis. The resulting exposure to
the sky is depicted in Fig.~\ref{exposure_map}.

The data were prepared following the procedures described in
\citet{jean03} and \citet{knoedlseder03}. SPI single detector events
have been gain corrected \citep[see][]{lonjou04b} and
binned into event spectra for each detector and instrument pointing
with 0.5~keV bins, leading to a 3-dimensional data space. In this
data space, the instrumental background in the 485--550~keV band was
modelled by two components: the instrumental 511~keV background line
and the underlying continuum background, as described in detail
in \citet{jean03}, \citet{teegarden04}, and \citet{jean04}.

The continuum background in the line analysis interval
508.5--513.5~keV was determined based on the continuum background
level at adjacent higher energies.
% {\bf TBD: details, time variation with GeDsat only?, etc.}.
%
The modelling of the instrumental line component has been
significantly improved since the first analyses by \citet{jean03} and
\citet{knoedlseder03}.
%As described in full detail in \citet{teegarden04, jean04} 
We employed a multi-component model for the instrumental 511~keV line,
which reflects to first order the multitude of prompt processes and
delayed (radioactive) decays that give rise to this background
line. We used e.g.\ the SPI background line
identifications \citep{weidenspointner03} and instrumental background
simulations employing the MGGPOD Monte Carlo suite
\citep{weidenspointner04a, weidenspointner04b} to identify the main
contributors to the 511~keV line. The prompt components were assumed
to scale with the rate of saturated Ge detector events
(GeDsat). Delayed components due to radioactive decays were modelled
by calculating a time history of their activity, assuming that their
production rate scales with GeDsat.

%%%%%%%%%%%%%%%
%   imaging   %
%%%%%%%%%%%%%%%
%
%\section{Imaging}
%\label{imaging}
%
%
%{\bf TBD}

%%%%%%%%%%%%%%%%%%%%%
%   model fitting   %
%%%%%%%%%%%%%%%%%%%%%
%
%\section{Model Fitting}
%\label{modelfitting}

%%%%%%%%%%%%%%%
%   results   %
%%%%%%%%%%%%%%%

\section{Results}
\label{results}

%A qualitative analysis of the 511~keV line emission morphology was
%performed by deconvolving the data into an intensity distribution on
%the sky using an implementation of the Richardson-Lucy algorithm
%\citep[see][ and references therein]{knoedlseder03} and a Bayesain
%approach \citep{allain04}. The results
%suggest an 

Imaging analyses of the 511~keV line emission with Richardson-Lucy
\citep[see][ and references therein]{knoedlseder03} and Bayesian
\citep{allain04} algorithms clearly show extended emission
from the Galactic bulge with a radial profile that is approximately
Gaussian. The sky maps do not provide any evidence for emission from
point sources, a PLE component, or (at this stage) the Galactic disk.

A more quantitative approach for studying the Galactic distribution
of the emission is model fitting. As a first step we have modelled the
bulge emission by a spherical distribution with a Gaussian radial
profile, located at the GC.
%Galactic center. 
We obtain a best fit FWHM of ${8^\circ}^{+3^\circ}_{-2^\circ}$, with
the quoted uncertainties defining the statistical $2\sigma$ confidence
limits. The corresponding bulge flux is $(0.96^{+0.21}_{-0.14}) \times
10^{-3}$~ph~cm$^{-2}$~s$^{-1}$, with the uncertainty being dominated
by the uncertainty of the width of the Gaussian intensity
distribution.  Spectroscopy of 511~keV line emission from the bulge
resulted in a best fit energy of $(511.02 ^{+0.08} _{-0.09})$~keV and
an intrinsic line width of $(2.67 ^{+0.30} _{-0.33})$~keV FWHM
\citep[see][]{lonjou04a}.

If we assume an ellipsoidal distribution with a Gaussian radial
profile we improve the fits marginally. A bulge distribution that is
more extended in longitude than latitude is slightly, but not yet
significantly, favoured. This elongation might be an indication of
emission from the Galactic disk.

\begin{figure}
\centering
\epsfig{file=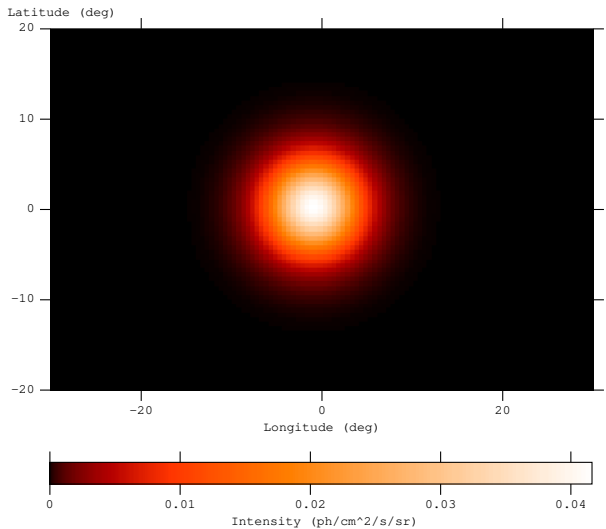,width=8cm,%
bbllx=80pt,bblly=198pt,bburx=535pt,bbury=598pt,clip=}
\caption{The best fit model for the 511~keV line emission from the
GC
%Galactic center 
region. A Gaussian, representing the Galactic bulge, is sufficient to
explain the data. Details are given in the text
\label{modmap}}
\end{figure}

If we do not constrain the spherical Gaussian model (FWHM
$8^\circ$) to be located at the GC,
%Galactic center, 
we find a best fit position of $l=-1.0^\circ
\pm 0.7^\circ$ and $b = 0.3^\circ \pm 0.7^\circ$ ($1\sigma$ error
bars). The best fit bulge model is depicted in Fig.~\ref{modmap}. The
significance of the displacement from the GC
%Galactic center 
is marginal at $1.6\sigma$.
% (see Fig.~\ref{confreg}). 
This displacement of the Gaussian slightly reduces the $2\sigma$ range
for FWHM, and slightly increases the bulge flux.

%\begin{figure}
%\centering
%\epsfig{file=gaussian-position_confidence-regions.ps,width=8cm,%
%bbllx=82pt,bblly=288pt,bburx=486pt,bbury=736pt,clip=}
%\caption{Confidence regions for the position of a Gaussian model (FWHM
%$8^\circ$) for the Galactic bulge emission. The best fit position is
%marked with an asterisk \label{confreg}}
%\end{figure}

%Adding a component for emission from the Galactic disk does not yet
%improve the fits significantly, but provides indications that
%such an emission exists. 
Although model fits yield hints of 511~keV line emission from the
Galactic disk, such a component has not yet been significantly
detected. The flux attributed to the disk depends strongly on the
assumed spatial distribution (in the following, we quote
2$\sigma$ upper limits). 
%The highest flux limit is found for a model
%assuming a 511~keV emissivity, constant in Galactic radius out to
%14~kpc, with an exponential scale height of 325~pc: $2.2 \times
%10^{-3}$~ph~cm$^{-2}$~s$^{-1}$. 
The DIRBE 35~$\mu$m and 240~$\mu$m maps, which trace old and young
stellar populations, yield flux limits of 1.4 and 0.9 $\times
10^{-3}$~ph~cm$^{-2}$~s$^{-1}$ when integrated over the Galaxy. The
motivation for chosing these two models was to span plausible positron
source distributions.
%These upper limits correspond to
%lower limits on B/D of 0.4, 0.5, and 0.8, respectively.

We tried to account for the 511~keV line emission with point sources
only using both model fitting and SPIROS \citep{skinner_connell03} imaging
analysis. A single point source can be excluded with a very high
confidence level. However, formally we cannot exclude the possibility
that the emission originates in a population of at least 4 point
sources.
We also investigated how much a single point source could contribute in
addition to our best (diffuse) description of the bulge emission. For
a point source at the position of Sgr~A$^\ast$, we can already set a
2$\sigma$ upper limit of about 40\% of the total flux.

Analyses of the putative PLE component are complicated by the fact
that the various OSSE studies did not converge on a unique
parametrization \citep[see table in][]{vonballmoos03}. In the
following, we restrict ourselves to a PLE model of \citet{purcell97},
a Gaussian at $l=-1.8^\circ$ and $b = 11.6^\circ$, with a FWHM of
$16.4^\circ$. Fitting this PLE model together with a Gaussian model
for the Galactic bulge has no significant effect on the results for
the latter. No significant flux is attributed to the PLE; a
conservative $2\sigma$ upper limit on the PLE flux (taking into
account the $2\sigma$ uncertainty in the bulge size) is $1.6 \times
10^{-4}$~ph~cm$^{-2}$~s$^{-1}$. It should be noted that a PLE of the
strength suggested by \citet{purcell97} would have already been
noticeable in the sky maps.

We have begun to investigate the positronium (Ps) continuum emission. First
results in the 409--505~keV band indicate that the sky distribution of
the total emission is consistent with what we find for the 511~keV
line.
If we use the \citet{harris90} result to fix the relative intensities
of Galactic diffuse and Ps continuum emission, we can infer a Ps
continuum flux from the measured total flux that is consistent with
previous findings.
However, because of systematic uncertainties in the subtraction
of Galactic continuum emission we refrain from quoting a value for the
Ps continuum flux and the Ps fraction until a forthcoming publication.

%%%%%%%%%%%%%%%%%%
%   discussion   %
%%%%%%%%%%%%%%%%%%

\section{Discussion}
\label{discussion}

With the available data, the sky distribution of the 511~keV positron
annihilation line emission from the GC
%Galactic center 
region as measured by SPI can be adequately described by a spherical
distribution with Gaussian radial profile and a FWHM of
${8^\circ}^{+3^\circ}_{-2^\circ}$, the quoted uncertainties being
statistical $2\sigma$ confidence limits. The bulge extent as
determined with SPI is consistent with the best fit OSSE results
($4^\circ - 6^\circ$) at the $2\sigma$ confidence level. We cannot
draw any firm conclusions regarding a possible discrepancy until we
have achieved a positive detection of the Galactic disk with SPI,
which might reduce the size of the bulge component.

We detect no significant deviation of the bulge distribution from
spherical symmetry. The data hint at a possible elongation in
longitude, which could reflect faint emission from the Galactic
disk. Allowing the location of the Gaussian bulge model to vary freely
suggests a slight offset from the GC
%Galactic center 
that is only marginally significant.
%results in a best fit position of $l=-1.0^\circ
%\pm 0.7^\circ$ and $b = 0.3^\circ \pm 0.7^\circ$ ($1\sigma$ error
%bars). The significance of the displacement from the Galactic center
%is marginal at $1.6\sigma$. 
However, it is interesting to note that our best fit position
coincides almost exactly with the centroid of the 511~keV line
emission 
%($l=-1.3^\circ \pm 0.75^\circ$ and $b = 0.3^\circ \pm 0.75^\circ$) 
%as determined 
found by \citet{tueller96} using OSSE data. If the existence of an
offset of the centroid of the 511~keV line emission can be
corroborated in future analyses including more data, this would set an
important constraint on emission models -- in particular on models that
attribute the bulk of the bulge 511~keV line emission to the
annihilation of light dark matter particles, whose distribution should be
exactly aligned with the dynamic center of the Galaxy. 
It is also interesting, but perhaps coincidental, that the centroid of
our best-fit Gaussian bulge distribution lies less than 1$\sigma$ away
from the hard X-ray source 1E1740.2-2942, which has been reported to
produce transient red-shifted annihilation line radiation \citep[see
review by][ and references therein]{harris97}.

Assuming that the centroid of the 511~keV line emission is the GC,
%Galactic center, 
we find a flux from the bulge of $(0.96^{+0.21}_{-0.14}) \times
10^{-3}$~ph~cm$^{-2}$~s$^{-1}$, with the uncertainty being dominated
by the uncertainty of the width of the Gaussian intensity
distribution. This value is in good agreement with previous results
obtained by high-resolution instruments 
\citep[see table in][]{jean03}, 
%which had small or moderate fields-of-view,
which had fields-of-view smaller than or comparable to that of SPI,
but falls short of the total 511~keV line fluxes obtained with OSSE
\citep[e.g.][]{milne00} and SMM \citep[e.g.][]{harris90} by a factor
of 2--3. SMM had a very large field-of-view ($\sim 130^\circ$ FWHM)
and therefore was sensitive to very extended emission beyond the
Galactic bulge.
%, including emission from the Galactic disk and possibly
%from a halo component. 
In case of OSSE, the flux was separated into bulge and disk
components, which had the values $(0.5-2.4)
\times 10^{-3}$~ph~cm$^{-2}$~s$^{-1}$ and $(0.8-2.6)
\times 10^{-3}$~ph~cm$^{-2}$~s$^{-1}$, respectively; these results strongly
depended on the assumed shape of the bulge component
\citep{milne00}. Our results for the bulge flux, and our $2\sigma$ upper
limits on the disk flux, $(0.9-2.2)
\times 10^{-3}$~ph~cm$^{-2}$~s$^{-1}$, are consistent with the OSSE
measurements.
% The SPI results
%for the bulge flux are consistent with those of OSSE
%is consistent with the O
%
%; the SPI value is consistent with the OSSE bulge measurement.
%Similarly, OSSE did detect emission from the
%Galactic disk that has not yet been significantly detected by
%SPI. The disk flux determined with OSSE covers a range of $(0.8-2.6)
%\times 10^{-3}$~ph~cm$^{-2}$~s$^{-1}$, but strongly
%depends on the assumed shape of the bulge component
%\citep{milne00}. Our $2\sigma$ upper limits on the disk flux, $(0.9-2.2)
%\times 10^{-3}$~ph~cm$^{-2}$~s$^{-1}$, are consistent with the OSSE
%measurements. 

Our upper limits on the flux from the Galactic disk derived from model
fits using the DIRBE 35~$\mu$m and 240~$\mu$m maps can be converted
into lower limits on B/D, which are 
%0.4, 
0.5 and 0.8, 
%the exponential model and 
respectively. Again, these limits are compatible with OSSE
measurements (0.2--3.3). The SPI limits in B/D already have
interesting implications. Based on their own low B/D, both DIRBE maps can
be excluded as sole source distribution of Galactic
positrons. Similarly, B/D for Galactic Type~Ia SNe \citep[from the
distributions by][]{dawson_johnson94,matteucci04} is below the SPI
limits. However, the observed distribution of low-mass X-ray binaries
is consistent with our results for B/D \citep{pranzos04}. If light
dark matter particles provide an important source of positrons, these
would have a high B/D value.

We do not find any evidence for a PLE component.

%%%%%%%%%%%%%%%%%%%%%%%
%   acknowledgments   %
%%%%%%%%%%%%%%%%%%%%%%%

\section*{Acknowledgments}

Based on observations with
INTEGRAL, an ESA project with instruments and science data centre
funded by ESA member states (especially the PI countries: Denmark,
France, Germany, Italy, Switzerland, Spain), Czech Republic and
Poland, and with the participation of Russia and the USA.

%%%%%%%%%%%%%%%%%%%%
%   bibliography   %
%%%%%%%%%%%%%%%%%%%%

% The following bibliography was produced with
\bibliographystyle{aa}
\bibliography{esapub}
% The results are inserted directly here to simplify
% the demonstration.

\end{document}